\documentclass[preprint]{aastex}

\shorttitle{Disk Formation by Dipole Fields}
\shortauthors{Matt et al.}

\slugcomment{Accepted by \apj.}

\begin{document}

\title{Disk Formation by AGB Winds in Dipole Magnetic Fields}

\author{Sean Matt and Bruce Balick}
\affil{Astronomy Department, University of Washington,
    Seattle WA 98195; \\ matt@astro.washington.edu,
    balick@astro.washington.edu}

\and

\author{Robert Winglee and Anthony Goodson}
\affil{Geophysics Department, University of Washington,
	Seattle WA 98195; \\ winglee@geophys.washington.edu,
	anthony@geophys.washington.edu}

\begin{abstract}

We present a simple, robust mechanism by which an isolated star can
produce an equatorial disk.  The mechanism requires that the star have
a simple dipole magnetic field on the surface and an isotropic wind
acceleration mechanism.  The wind couples to the field, stretching it
until the field lines become mostly radial and oppositely directed
above and below the magnetic equator, as occurs in the solar wind.
The interaction between the wind plasma and magnetic field near the
star produces a steady outflow in which magnetic forces direct plasma
toward the equator, constructing a disk.  In the context of a slow (10
km s$^{-1}$) outflow ($10^{-5}$ M$_\odot$ yr$^{-1}$) from an AGB star,
MHD simulations demonstrate that a dense equatorial disk will be
produced for dipole field strengths of only a few Gauss on the surface
of the star.  A disk formed by this model can be dynamically important
for the shaping of Planetary Nebulae.

\end{abstract}

\keywords{MHD --- planetary nebulae: general --- stars: AGB and
post-AGB --- stars: magnetic fields --- stars: winds, outflows}

\section{Introduction}

Asymmetry presents one of the greatest challenges to our understanding
of stellar outflows.  Planetary nebulae (PNe) comprise a fairly well
studied set of objects of which more than 80\% are axisymmetric
\citep{za86,b87,sea93}.  Although the statistics for proto-PNe are
limited \citep{st98}, the morphological statistics seen in PNe appear
to apply.  For this reason, and because they are generally bright and
easily studied observationally, PNe are an attractive set of objects
to study if we are to gain insight into the physical mechanisms
producing aspherical outflows.

In the past, the properties of PNe have been explained by an
interacting stellar winds model \citep{kea78,kw85,b87}.  This holds
that a fast, tenuous wind ($v\sim$1000 km s$^{-1}$, $\dot{\rm M} \sim
10^{-7}$ M$_\odot$ yr$^{-1}$) from a PNe central star encounters an
older, slower wind ($v\sim$10 km s$^{-1}$, M $\leq 0.1$ M$_\odot$).
The slow material was expelled earlier ($\dot{\rm M} \sim 10^{-5}$
M$_\odot$ yr$^{-1}$), while the central star was on the asymptotic
giant branch (AGB) or in the post-AGB phase.  The shocked region
between the two winds produces an expanding, wind-blown bubble (WBB)
that is visible due to radiative cooling.  To produce axisymmetric
WBB's, the model assumes that the slow wind is aspherically
distributed so that it is densest in the equatorial plane.  Direct
observational evidence for dynamically important disks, tori, and
their remnants in the environment of PNe is clear at visible,
infrared, and radio wavelengths
\citep{hug00,mei00,b87}.

The central astronomical issue is how and why the slow winds can be
equatorially concentrated.  Physically, what processes are at work to
build a disk?  Winds from normal AGB and post-AGB stars are expected
to be largely spherical, since the stellar surface is spherical and
the wind driving force, radiation \citep{sea97,hea94}, is isotropic.
Thus, as the star evolves from the AGB into a white dwarf, the wind
geometry is expected to remain symmetric.

However, as noted above, most PNe and proto-PNe show strong
axisymmetries.  Proposed mechanisms that might produce outflowing
disks (see, e.g., Frank 1999 for a review) require reserves of angular
momentum far in excess of that in the sun \citep{s97}.  While some PNe
may have formed from stars with companions close enough to transfer
orbital angular momentum into the outer layers of a fully distended
AGB star \citep{s98a,livio82}, a mechanism that accounts for the
nearly ubiquitous incidence of aspherical PNe is necessary.  In
addition, bipolar nebulae have a low galactic scale height and,
consequently, are believed to evolve from relatively massive
progenitors \citep{cs95,zg88}.  If so, a mechanism that produces
stronger collimation exclusively around massive stars is needed.

Recently, several authors have discussed magnetic shaping mechanisms.
\citet{cl94} developed an analytical model in which a toroidal
magnetic field carried in a fast wind produces prolate and bipolar WBB
morphologies, even if the slow wind is spherically symmetric.
\citet{rf96}, \citet{gs97}, and \citet{gsea99} confirmed this model
via numerical magnetohydrodynamic (MHD) simulations, demonstrating
that toroidal magnetic fields carried in fast winds can constrain WBB
outflow along the equator, producing a wide variety of polar flows.
Their magnetic field topology is derived from what is expected in the
solar wind, where the poloidal field lines connecting the outflowing
wind to the solar surface are wound up due to the rotation of the sun.
However, these and other \citep{p92,pea92} models have only considered
the effect of toroidal magnetic fields.  The influence of {\it
poloidal} magnetic fields on AGB winds is yet unexplored.

The sun has poloidal magnetic field that is dynamically important to
the solar wind, as evidenced by streamers and coronal mass ejections.
In fact, the sun's magnetic field is dominated by the (global) dipole
component outside a few solar radii throughout most of the magnetic
cycle \citep{bea98}.  Several authors
\citep{mestel68,pk71,washimiea87,ms87,washimi90,ws93,banaszkiewiczea98,
kg99,kg00} have shown that the dipole field has a significant
influence on the solar wind.  \citet{washimiea87} and \citet{kg99}
reported an equatorial density enhancement in the solar wind, due to
the poloidal field (in the absence of rotation).  It therefore may be
insufficient to consider the effects of toroidal magnetic fields alone
on stellar winds without including the effects of the poloidal fields.

Our main goal is to show that a simple poloidal magnetic field, a
dipole, can be dynamically important in stellar winds.  We propose and
justify a simple mechanism that might form strongly axisymmetric winds
in isolated AGB or post-AGB stars without the need for large amounts
of rotation.  In the presence of a dipole field, wind plasma is
deflected toward the equator, and a disk can be produced.  We detail
the qualitative aspects of the model in \S \ref{model} and present
numerical magnetohydrodynamic (MHD) simulations that confirm the
conceptual model in \S \ref{simulations}.  A discussion of some of the
implications of this work is included in \S\ref{discussion}.

\section{The Conceptual Model \label{model}}

Consider a wind from a spherical star.  In the absence of rotation and
magnetic fields, the launching of the wind is completely isotropic if
the wind is driven by radiation pressure on dust or thermal pressure.
The wind accelerates to a terminal velocity within several stellar
radii \citep{lc99}.  However, if the star initially has a dipole
magnetic field on the surface, we demonstrate that the end result is
an outflow that is densest in the magnetic equatorial plane.

\begin{figure}[t]
\plotone{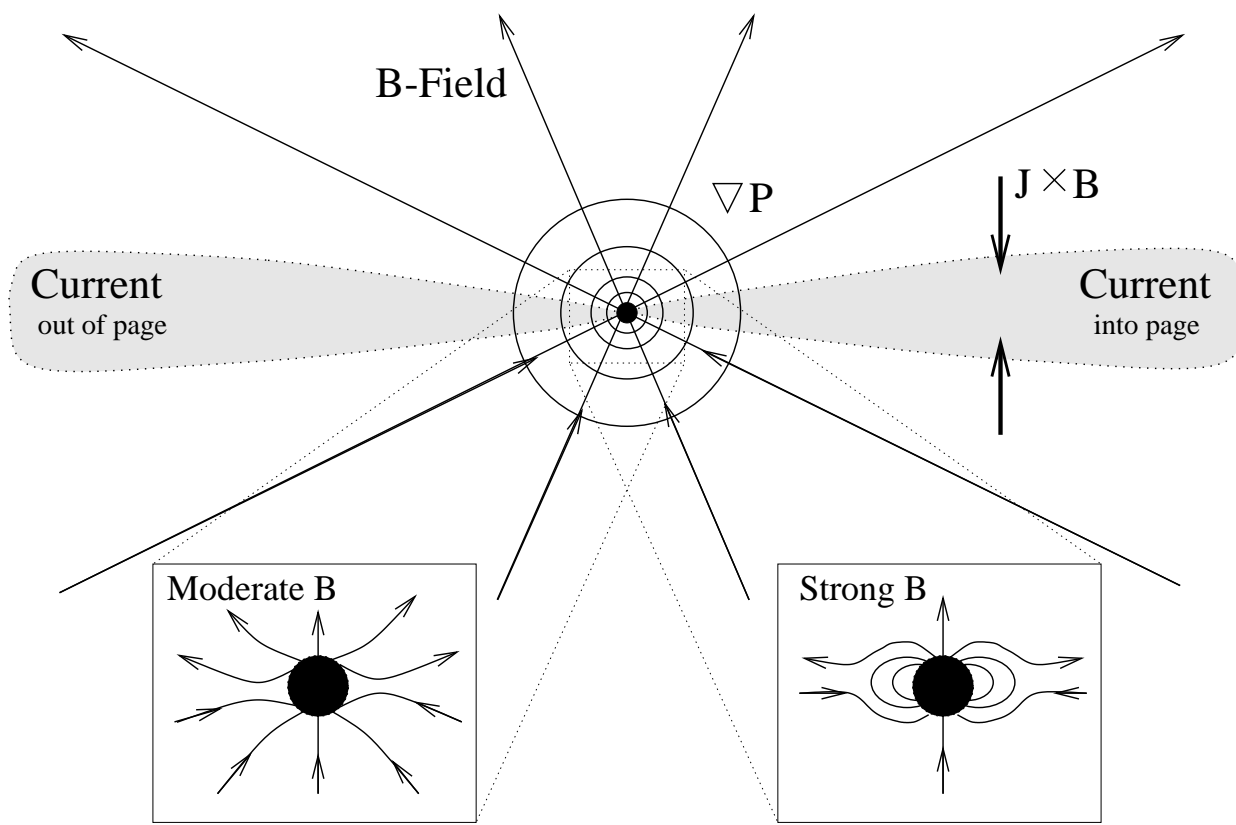}

\caption{A qualitative cartoon of the model including magnetic field
lines and pressure contours is shown.  In steady-state, a weak,
initially dipolar magnetic field ($\mbox{\boldmath $B$}$) becomes
radial but oppositely directed above and below the magnetic equator,
and a wind is driven isotropically from the surface of the star.  An
equatorial current sheet ($\mbox{\boldmath $J$}$) exists to maintain
the radial magnetic field.  The resulting ($\mbox{\boldmath $J$}\times
\mbox{\boldmath $B$}$) force is directed toward the
equator.  For moderate (left inset) or strong (right inset) dipole
magnetic fields, the magnetic force is sufficient to divert the wind
toward the equator.  \label{model_fig}}

\end{figure}

For this to occur, the wind must be sufficiently ionized so that the
magnetic field is frozen into the plasma.  If the total (thermal plus
kinetic) energy of the wind is much greater than the magnetic energy,
the magnetic field is carried radially outward in the wind.  The
magnetic field topology becomes radial everywhere.  The field
direction is inherited from the dipole in that the field lines are
oppositely directed above and below the equator.  This can be
understood if one considers one end of each field line to be
``anchored'' to the surface of the star while the other end is
stretched radially outward in the wind.  In this case (very weak
field), the wind will be isotropic.  Figure \ref{model_fig}
illustrates the approximate topology of a weak magnetic field.  In
order to maintain a radial magnetic field (with an equatorial
direction reversal), an azimuthally directed current sheet must exist
in the equatorial plane.  As shown in Figure \ref{model_fig}, the
resulting magnetic force (${\mbox{\boldmath $J$}} \times
\mbox{\boldmath $B$}$, where \mbox{\boldmath $J$} is the volume
current and \mbox{{\boldmath $B$}} the magnetic field) is directed
toward the equator.  

If the magnetic field energy on the surface of the star is comparable
to or greater than the energy in the wind, this ${\mbox{\boldmath
$J$}} \times \mbox{\boldmath $B$}$ force will be strong enough to
redirect the wind.  In other words, plasma will be forced to flow
along the dipole field lines near the star.  Since the plasma is
frozen-in to the magnetic field, continuity requires that the wind
momentum is proportional to $r^{-3}$, since $\mbox{\boldmath $B$}
\propto r^{-3}$ for a dipole.  For an accelerating wind, the kinetic
energy will therefore decrease more slowly than $r^{-3}$, while the
dipole magnetic energy decreases as $r^{-6}$.  Hence, the magnetic
field is only effective near the star.  So no matter how strong the
field is, there will always be an exterior region of the outflow that
is dominated by the kinetic energy of the wind (and where the field
lines become radial).  Numerical simulations presented in \S
\ref{simresult} and done by others \citep{pk71,ws93,kg99,kg00} confirm
this idea.

For very strong magnetic fields, the wind may be completely quenched
near the equator \citep{kg00}.  In this case the low-latitude magnetic
field lines remain closed (each end attached to the star at equal but
opposite latitudes, forming a static loop), and no ionized particles
can flow from this region (ignoring diffusive effects).  However, the
high-latitude magnetic field lines would still open (tending toward
radial at large radii), but the plasma leaving at these high latitudes
will first be forced toward the equator \citep{pk71,washimiea87,kg99}.

The two insets in Figure \ref{model_fig} show the approximate magnetic
field topology in the wind near a star with a moderate (magnetic and
wind energy comparable) and strong (magnetic energy dominates) initial
dipole magnetic field.  As shown, plasma flowing along open field
lines is first directed more toward the equator than radial lines and
eventually tend toward radial.  The net effect of this flow pattern is
an increase of the wind density and thermal pressure near the equator,
and a decrease near the poles.  This is similar to what happens in
solar coronal streamers.  Open streamers have all open field lines,
helmet streamers have some closed loops, and both types show density
enhancements along radial lines through their centers.  The model
presented here can superficially be thought of as a star with an
axisymmetric, global streamer encircling its equator, thus forming a
disk.  It is this concept that we propose may be responsible for
forming outflowing disks from AGB or post-AGB stars.

In addition to having a density anisotropy, the redistribution of gas
pressure can affect, via $\mbox{\boldmath $\nabla$} P$ forces, the
radial force on the wind such that the wind on the equator is
accelerated more slowly than the wind on the poles.  This leads to a
wind that is slowest near the equator.  In many astrophysical winds,
gas pressure gradients are negligible, and so the wind velocity may be
unaffected by this.  Such winds, however, would still be deflected
toward the equator as discussed above.

It is instructive to note that the strength of a dipole magnetic field
falls off as $r^{-3}$, while a radial field goes as $r^{-2}$.  This
means that there is more total magnetic energy in the radial field
than in the dipole field.  This energy is added by the wind as it does
work to stretch out the field lines.  In other words, no magnetic flux
is carried off of the surface of the star (once the wind is
established), though the stretched field remains anchored on the star,
and the axisymmetric density distribution remains intact.  This is
because the $\mbox{\boldmath $J$} \times \mbox{\boldmath $B$}$ forces
balance the gas and dynamic pressure gradient forces, and the plasma
flow is parallel to the magnetic field.  Hence, the star does not need
to continue to produce any new magnetic energy (after the initial
dipole is produced), provided the diffusion timescale is significantly
longer than the dynamical timescale of the outflow.  However,
observations of PNe progenitors and central stars show that the wind
speed increases from $\sim 10$ km s$^{-1}$ to $\sim 1000$ km s$^{-1}$
as the star evolves while $\dot{\rm M}$ decreases from $\sim 10^{-4}
\; {\rm M}_\odot$ yr$^{-1}$ to $\sim 10^{-8} \; {\rm M}_\odot$
yr$^{-1}$ \citep{f99}.  The stellar surface temperature, radius,
gravity, and luminosity all evolve rapidly.  It is therefore plausible
that the magnetic field is only of transient importance in structuring
the wind geometry of these stars.

\section{MHD Simulations \label{simulations}}

Here we present numerical MHD simulations testing this model in the
context of winds from the surface of an AGB star.  The simulations are
simplified in that they do not include the effects of rotation,
gravity, or radiative cooling, they are carried out in two dimensions,
and the wind is driven radially by thermal pressure at the star's
surface.  They are meant to serve only as a demonstration of
feasibility for the MHD model, and to investigate the types of results
that can be expected for more sophisticated computational models.

We have neglected rotation to study winds from isolated stars (that
is, they are not spun-up).  A giant star with solar angular momentum
is expected to rotate very slowly relative to its breakup velocity
\citep{pascoli87,s97}.  Presumably, a star must have some amount of
rotation in order to generate a magnetic field, via a stellar dynamo
\citep{parker79}, but if the rotation velocity on the surface is much 
less than the outflow velocity of the wind, we assume that rotational
influence on the isotropy of the wind is negligible.

A more realistic treatment would include the effects of gravity.  This
would affect the details of the acceleration of the wind, but would
not change the results of the qualitative model.  The simulations
presented here are most applicable to stars in which the radiation
pressure balances gravity above the surface so that only pressure and
magnetic forces are important.  We want to stress, however, that the
model is not dependent on the wind driving mechanism.

Below, we discuss the code and boundary conditions (\S \ref{simcode}),
initial conditions (\S \ref{siminit}), and simulation results for
various magnetic field strengths (\S \ref{simresult}).

\subsection{Simulation Code \label{simcode}}

The simulation code uses a two-step Lax-Wendroff (finite difference)
scheme \citep{rm67} to simultaneously solve the single-fluid, ideal MHD
equations
\begin{eqnarray}
{{\partial\rho}\over{\partial t}} &=& 
	-\mbox{\boldmath $\nabla$}\cdot(\rho \mbox{\boldmath $v$}) \\
{{\partial(\rho \mbox{\boldmath $v$})}\over\partial t} &=& 
	-\mbox{\boldmath $\nabla$} (\rho\mbox{\boldmath $v$}^2 + P) 
	+{{1}\over {\rm c}}(\mbox{\boldmath $J$}\times\mbox{\boldmath $B$})  \\
{{\partial e }\over\partial t} &=& 
	-\mbox{\boldmath $\nabla$}\cdot[\mbox{\boldmath $v$}(e + P)]
	+\mbox{\boldmath $J$} \cdot \mbox{\boldmath $E$} \\
{{\partial\mbox{\boldmath $B$}}\over\partial t} &=& 
	-{\rm c}(\mbox{\boldmath $\nabla$}\times\mbox{\boldmath $E$})
\end{eqnarray}
in two dimensions, and uses
\begin{eqnarray}
\mbox{\boldmath $E$} &=& 
  -{{1}\over{\rm c}}(\mbox{\boldmath $v$}\times\mbox{\boldmath $B$}) \\
\mbox{\boldmath $J$} &=& 
  {{\rm c}\over {4\pi}}(\mbox{\boldmath $\nabla$}\times\mbox{\boldmath $B$}) \\
e 	&=& 
  {{1}\over {2}}\rho\mbox{\boldmath $v$}^2 + {{P}\over {\gamma - 1}}
\end{eqnarray}
where $\rho$ is the density, \mbox{\boldmath $v$} the velocity, $P$
the scalar gas pressure, $e$ the internal energy density,
\mbox{\boldmath $B$} the magnetic field, \mbox{\boldmath $J$} the 
volume current, \mbox{\boldmath $E$} the electric field, c the speed
of light, and $\gamma$ the ratio of specific heats (we used $\gamma$ =
1.4 in our simulations).

We employ a three-layer boundary condition at the base of the wind to
avoid complications introduced by the finite-difference scheme that
numerically violate $\mbox{\boldmath $\nabla$} \cdot \mbox{\boldmath
$B$} = 0$ near fixed boundaries.  On the innermost boundary, all
plasma quantities ($P$, $\rho$, $\mbox{\boldmath $v$}$) are constant
(in time) and $\mbox{\boldmath $B$}$ is held at the dipole value.  The
intermediate boundary is the same as the inner boundary except that
$\mbox{\boldmath $B$}$ is time-dependent.  The outer boundary is the
same as the inner boundary except that both $\mbox{\boldmath $B$}$ and
$\mbox{\boldmath $v$}$ are time-dependent.  We consider the
intermediate boundary to be the ``surface'' of the star, though the
density and pressure are constant on the outer-most layer.  These
boundary conditions allow the magnetic field on the stellar surface to
be carried outward in the wind in a manner that preserves
$\mbox{\boldmath $\nabla$} \cdot \mbox{\boldmath $B$} = 0$.  Outside
the star, all quantities are time-dependent.  The large-scale results
are not sensitive to the details of these boundary conditions.

The simulations use a $500 \times 500$ point Cartesian grid and a
``star'' with a radius of six gridpoints at the center.  If we assume
the star has a radius of $R_* = 1.5 \times 10^{13}$ cm, this
corresponds to a total ``field of view'' of $83 \times 83$ AU with a
``resolution'' of 1/6 AU.  The calculations are strictly 2-D, and no
symmetry constraint is imposed.

The 2-D dipole equation is given in polar coordinates as
\begin{eqnarray}
\label{dipole_eqn}
B_r      &=& M \left({R_*\over r}\right)^2 \cos\theta  \nonumber \\
B_\theta &=& M \left({R_*\over r}\right)^2 \sin\theta
\end{eqnarray}
where $M$ is the field strength at the surface, $R_*$ is the radius of
the object, and $\theta$ is the angle from the magnetic pole.  This
field satisfies $\mbox{\boldmath $\nabla$} \cdot \mbox{\boldmath $B$}
= 0$ and $\mbox{\boldmath $\nabla$} \times \mbox{\boldmath $B$} = 0$
(the dipole is current-free) in two dimensions.  It is different from
a 3-D dipole in that the field strength at the poles is equal to
(instead of twice) that at the equator, and the 2-D dipole falls
off as $r^{-2}$ (instead of $r^{-3}$).

\subsection{Initial Conditions \label{siminit}}

While these simulations apply to a family of systems (i.e.\ they are
scalable), we report numerical parameters and results for a system in
which the star has a radius of 1 AU.  If the physical values of
$\mbox{\boldmath $v$}$, $\rho$, $P$, and $\mbox{\boldmath $B$}$ remain
unchanged, the results apply to a star of arbitrary radius, distances
and times scale with $R_*$, and $\dot{\rm M}$ with $R_*^2$.  In this
way, we can superficially consider outflow from an AGB and post-AGB
star simultaneously.

Before the simulations begin, the star has a constant density of $n =
3.4 \times 10^{11}$ cm$^{-3}$ and a pressure such that the sound speed
is 5.9 km s$^{-1}$ on its three-layer boundary.  These conditions lead to a
mass flux and outflow velocity (see \S \ref{simresult}) that are
reasonable for a wind from a cool AGB star.  The star also has a
dipole magnetic field (Eqn. \ref{dipole_eqn}) initialized everywhere
in the simulation region.  The velocity of the plasma is initially
zero everywhere.

When the simulation begins, the difference between the stellar surface
pressure and that of the surrounding ambient material leads to an
outward acceleration of gas, which plows the ambient material ahead of
it.  After the ambient gas is completely swept off of the simulation
grid, a steady-state outflow is established whose properties only
depend on the conditions on the surface of the star.  In fact,
provided that the outflow becomes supersonic, the results are
insensitive to the initial conditions of the ambient gas.  Since we
were only interested in the characteristics of the steady-state
outflow, we chose the density and sound speed of the ambient gas ($1.0
\times 10^9$ cm$^{-3}$ and 11.8 km s$^{-1}$, respectively) so that it is
efficiently cleared out of the simulation region.  It should be noted
that the steady state of the outflow is due to the unchanging (in
time) boundary conditions on the surface of the star.  On real stars,
dynamic surface processes and stellar evolution require that winds are
time-dependent (on long enough timescales).  We are, however, only
currently interested in determining the properties of the wind while
the star satisfies the conditions outlined above.

In order to determine whether the magnetic field might influence the
wind, one can use the ratio of the wind energy density to the magnetic
energy density.  This ratio increases with distance from the star (see
\S \ref{model}), so the magnetic field will be most effective near the
surface.  If the wind has a constant energy from the stellar surface
outward, this ratio on the surface ($r = 0$) is given by $\dot{\rm M}
v_{\rm f} R_*^{-2} \mbox{\boldmath $B$}_0^{-2}$, assuming that thermal
pressure is negligible far from the star and where $v_{\rm f}$ is the
terminal velocity of the wind.  If energy is added to the wind above
the surface (e.g., via radiation pressure or magnetic effects), this
is an upper limit.

This formula for the ratio of wind to magnetic energy is convenient
because it gives the relative importance of a hypothetical or measured
magnetic field using measured wind parameters.  By setting it equal to
unity, one obtains (roughly) a minimum magnetic field strength capable
of deflecting the stellar wind.  Note that this formula determines
{\it whether} the magnetic field is important, while the topology of
the field determines {\it how} the wind is affected.  For an AGB star
with $\dot{\rm M} \sim 10^{-5} \; {\rm M}_\odot {\rm yr}^{-1}$,
$v_{\rm f} \sim 10$ km s$^{-1}$, and $R_* \sim 1.5 \times 10^{13}$ cm,
this implies that the wind and magnetic energy densities are
comparable for a magnetic field strength of 1.7 Gauss on the surface.
For this reason, we ran seven different simulations with surface
magnetic field strengths bracketing this value.

In the simulations, the outflow is pressure-driven, so the total
energy in the wind as it accelerates outward is equal to its thermal
energy on the surface (where $\mbox{\boldmath $v$} = 0$).  For the
simulations, then, the key parameter (the ratio of wind to magnetic
energy densities) is simply
\begin{equation}
\label{beta_eqn}
\beta = {{8 \pi P_0} \over {\mbox{\boldmath $B$}_0^2}}
      =	{{\rm thermal~ energy}\over{\rm magnetic~ energy}}
\end{equation}
where the subscript 0 denotes quantities initially on the stellar
surface.  In all of the simulations, the initial gas properties are
constant and only the strength of the magnetic field varies.

\subsection{Simulation Results \label{simresult}}

The simulations comprise seven different cases spanning $\beta =
\infty$ to 0.1, corresponding to $\mbox{\boldmath $B$}_0 = 0$ and 6.0
Gauss on the surface of the star, respectively.  The first two columns
of Table \ref{results_tab} list the values of $\beta$ for all cases
and the corresponding dipole magnetic field strength initially on the
surface of the star.

\begin{figure}[t]
\plotone{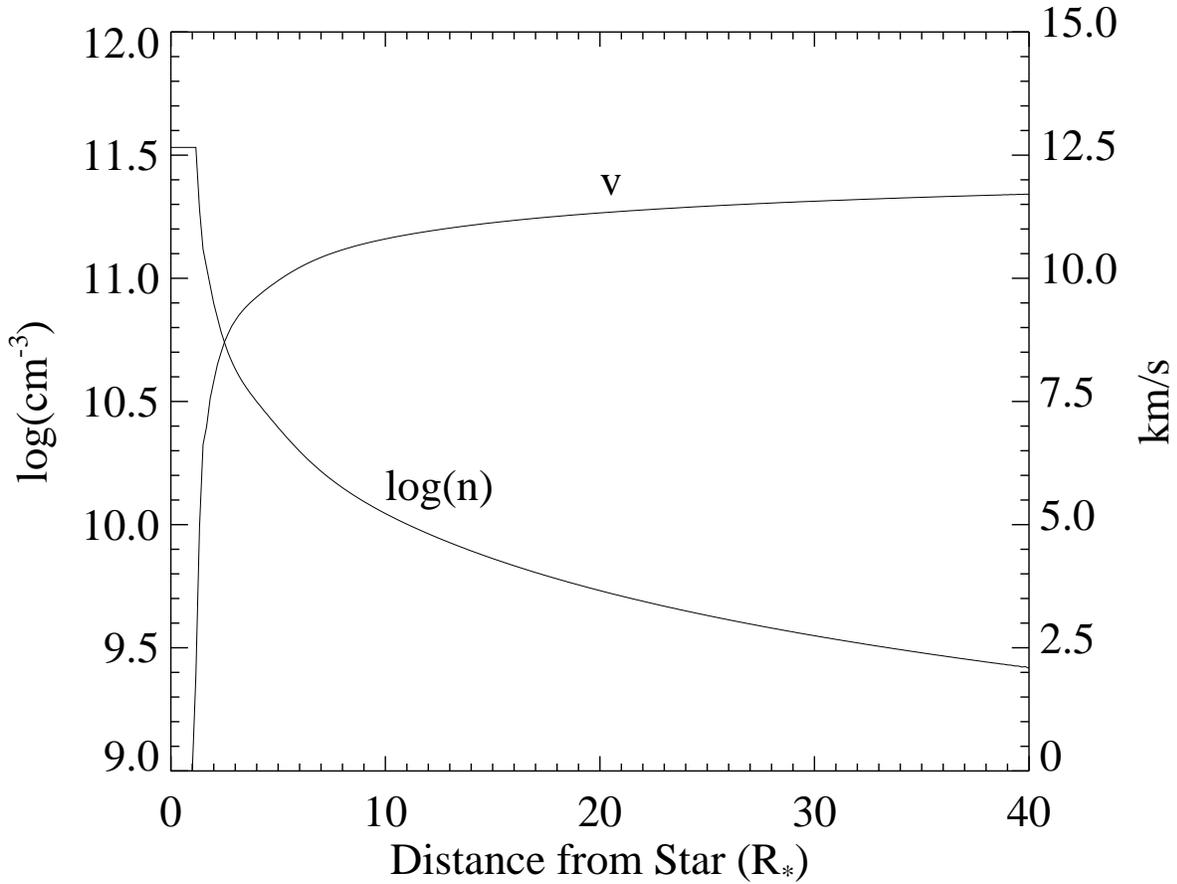}

\caption{The isotropic wind density and velocity as a function of
distance from the star is shown for the steady-state, non-magnetic
outflow.
\label{profile_fig}}

\end{figure}

The $\beta = \infty$ case is mainly used as a reference to which we
compare the cases that include magnetic fields.  Figure
\ref{profile_fig} shows the density and velocity profile of this
non-magnetic case in steady-state.  The wind accelerates to $\sim 11$
km s$^{-1}$ within 10$R_*$ (becoming supersonic) and increases very
little as it travels further from the star.  Using the density and
velocity just outside the surface of the star, the calculated 2-D mass
flux is $2.8 \times 10^{-6}$ M$_\odot$ yr$^{-1}$ AU$^{-1}$ ($ =
2\pi\rho\mbox{\boldmath $v$}r$) for pure hydrogen.  This quantity is
the mass flux per length normal to the 2-D plane.  The equivalent mass
flux for a 3-D system would be $\dot{\rm M} = 5.6 \times 10^{-6}$
M$_\odot$ yr$^{-1}$, assuming the star is spherical.  The velocity and
mass flux of this outflow are consistent with assumed properties in
the winds of AGB stars \citep{f99}.

All of the cases established steady-state outflows in the entire
simulation region in less than 200 years, comparable to the time
needed to clear out the initial ambient gas.  Figure
\ref{datamodel_fig} shows the magnetic field lines and gas pressure
contours overplotting a grey-scale image of the current density for
the case with $\beta = 0.2$.  A comparison with Figure \ref{model_fig}
indicates a confirmation of the qualitative model (\S \ref{model}).
The current and magnetic field lines in the simulation are oriented in
the same direction as predicted by the model.  The pressure gradient
is significantly anisotropic such that it is steeper on the poles than
on the equator (near the star), as expected.

\begin{figure}[t]
\plotone{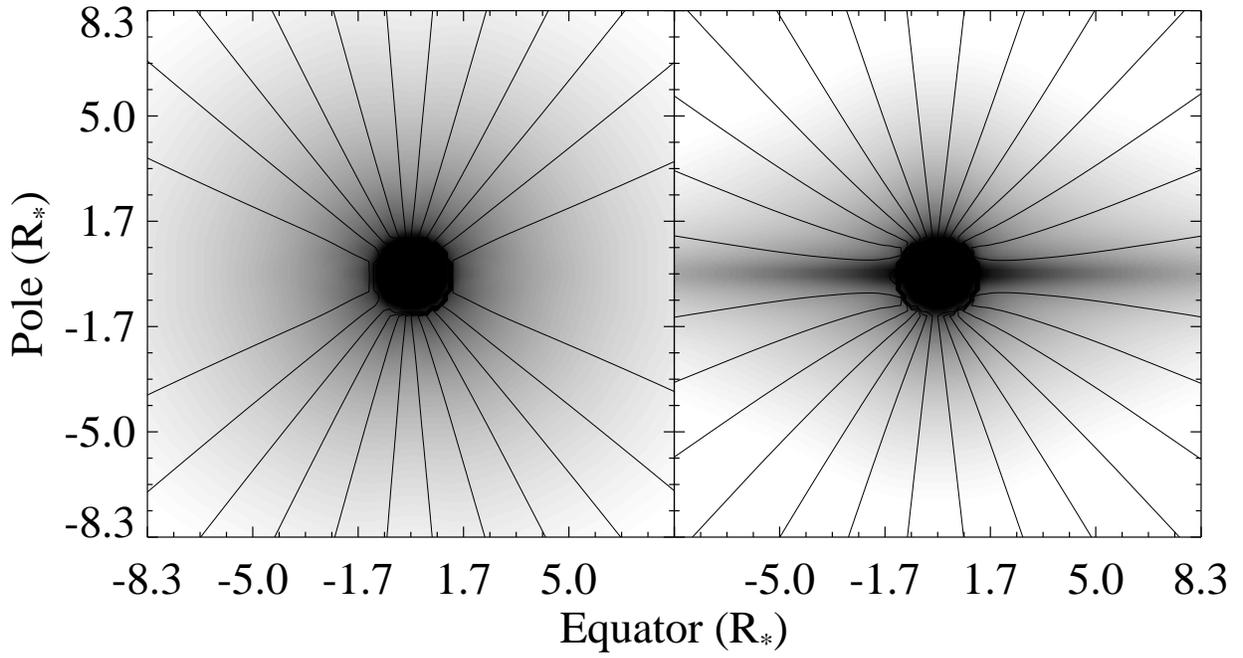}

\caption{A grey-scale image of $\log n$ with magnetic field 
lines overplotted is shown for the $\beta =$ 5.0 (left) and $\beta =$
0.1 (right) cases in steady-state.  Only the inner $100 \times 100$
grid points are shown.  White corresponds to $n \le 10^{10}$ cm$^{-3}$
while $n \ge 10^{11.5}$ cm$^{-3}$ is black.  The line spacing is
proportional to field strength. \label{densfld_fig}}

\end{figure}

An interesting feature of Figure \ref{datamodel_fig} is that the
magnetic field lines are not perfectly radial.  Far from the star, the
lines tend toward radial, but near the star, they are directed more
toward the equator than radial lines.  This result is expected since,
in this case, the magnetic energy is greater than the wind energy on
the stellar surface (see \S \ref{model}).  The dependence of magnetic
field topology on field strength is even more apparent in Figure
\ref{densfld_fig}, which shows the density and field lines for the
$\beta = 5.0$ and $0.1$ cases.  The lines are almost completely radial
in the weak field case (left panel), while for the stronger field
(right panel), they retain much of their original (dipole) structure
near the star.  A comparison of Figures \ref{densfld_fig} and
\ref{model_fig} suggests that our simulation with the strongest magnetic
field should still be considered to have a ``moderate'' field
strength, since their are no large closed field lines.  Since the
flows are in steady-state, the magnetic field lines are in the same
direction as streamlines for wind particles.  With this in mind, it is
easier to understand the equatorial density enhancement visible in the
right panel of Figure
\ref{densfld_fig}.

\begin{figure}
\plotone{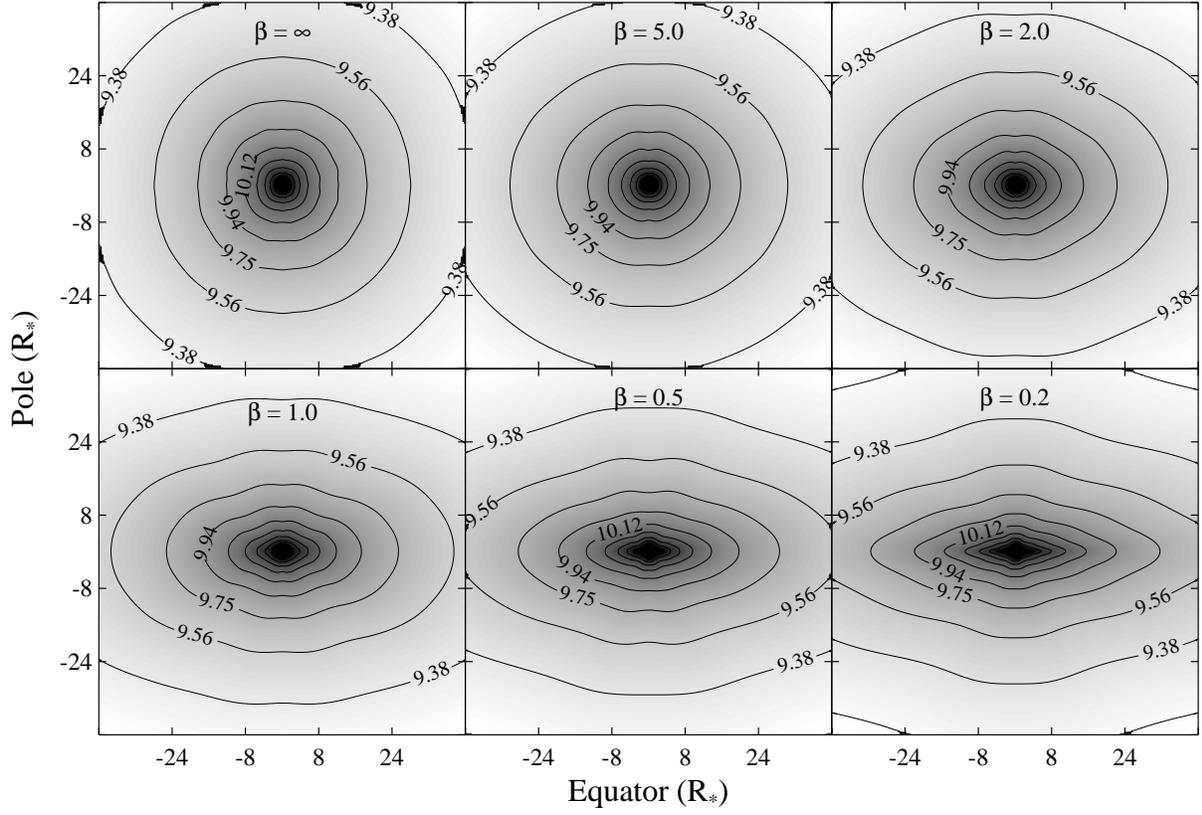}

\caption{Maps of $\log n$ (cm$^{-3}$, grey-scale and contours) for
various cases reveal a relationship between the magnetic field
strength and the density distribution of the wind.  The value of
$\beta$ is indicated at the top of each panel, and the entire
simulation region is shown in its steady-state configuration.
\label{denscont_fig}}

\end{figure}

Figure \ref{denscont_fig} shows the density distribution in the
outflow for six cases.  As $\beta$ decreases, the density contours
become less circular, having a major axis along the equator and a
minor axis along the poles.  This indicates a correlation between the
equatorial concentration of the wind and the magnetic field strength.
For the weak-field cases ($\beta > 1$, top row of Fig.\
\ref{denscont_fig}), the density contours are only slightly
different (by a few tenths or less) from circles.  For smaller
$\beta$, however, the magnetic field has a stronger influence on the
outflow properties, and the density contours are much more disk-like.

\begin{figure}[t]
\plotone{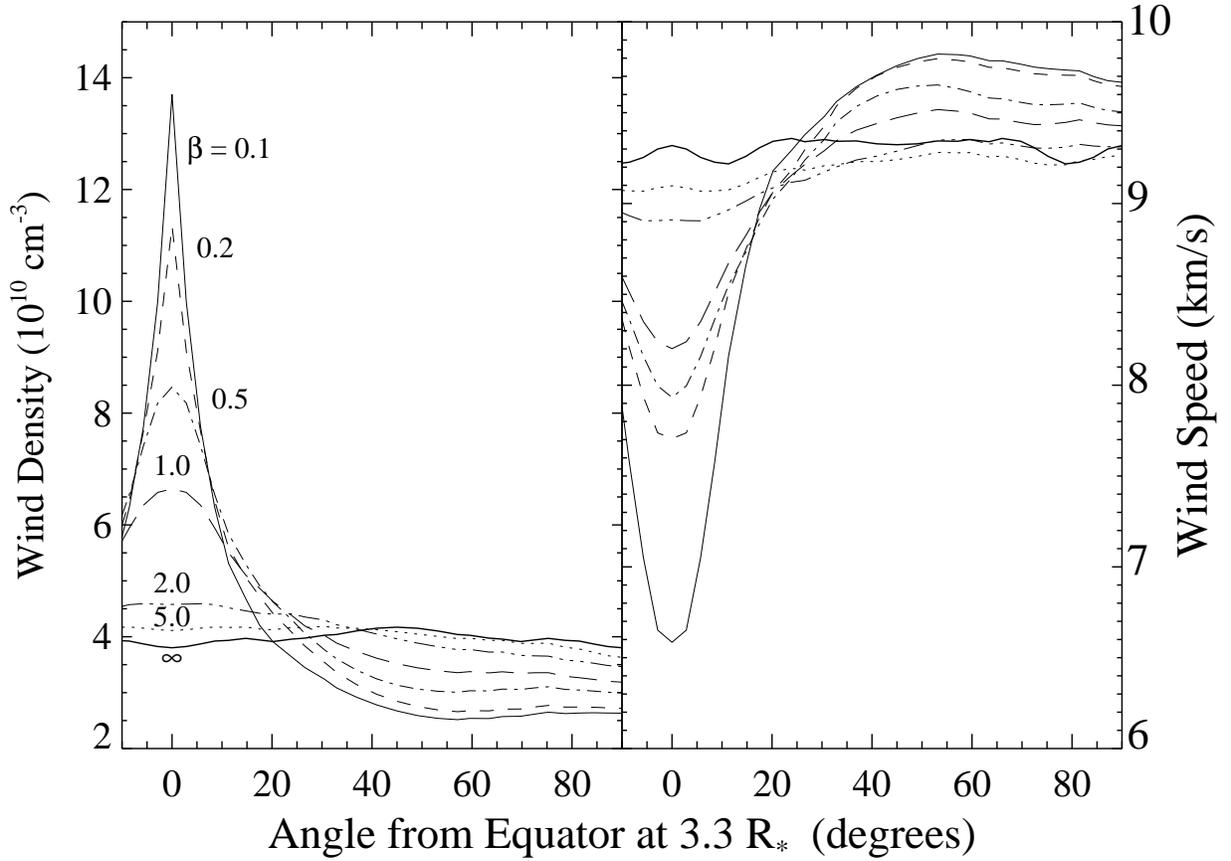}

\caption{Shown is the density (left panel) and wind speed (right panel)
at 3.3$R_*$ as a function of angle from the magnetic equator for all
seven cases.  In the left panel, the value of $\beta$ is indicated
next to the peak density of each case.  In the right panel, each line
style corresponds to the same case as indicated in the left panel.
\label{angqtys_fig}}

\end{figure}

This trend is more apparent in the left panel of Figure
\ref{angqtys_fig}, a plot of density as a function of latitude at
$3.3R_*$ for all cases.  The non-magnetic case (thick line) is
isotropic, but for decreasing $\beta$, the outflow is denser near the
equator and more tenuous near the poles.  This illustrates that the
wind plasma at high latitudes is moved toward lower latitudes by
magnetic forces.  For $\beta \le 1.0$, steep density profiles indicate
the presence of an outflowing disk that is fed by high latitude
material.  For the $\beta = 0.1$ and 0.2 cases, the flatness of the
disk is unresolved on the simulation grid near the star.  This is
evident in the left panel of Figure \ref{angqtys_fig}, as these two
density profiles display a sharp point (cusp) at the equator.  A
velocity trend is evident in the right panel of Figure
\ref{angqtys_fig} where the wind speed is shown as a function of
latitude at a radius of $3.3R_*$.  Here, it is evident that the wind
is slower near the equator and faster near the poles for decreasing
$\beta$.

The outflow in the non-magnetic case deviates slightly from purely
isotropic due to grid orientation effects that arise from placing a
round star in a Cartesian grid, coupled with difficulties of the
numerical scheme on the ``stair-step'' corners of the gridded stellar
surface.  This is apparent in the upper left panel of Figure
\ref{denscont_fig} and the thick line in both panels of Figure
\ref{angqtys_fig}.  The very slight anisotropy of the non-magnetic 
outflow represents an uncertainty in the results for all cases, due to
these gridding effects.  At any radius, the outflow properties
($\rho$, $\mbox{\boldmath $v$}$, $P$) of the non-magnetic case vary by
less than 10\%.  It is also important to note that the outflow
properties on the poles of the non-magnetic flow are identical to
those on the equator.

The equator-to-pole density ratio, $q$ (the ``density contrast''), is
a useful way to characterize the axisymmetric inertial properties of
the wind.  Figure \ref{contdens_fig} shows the density contrast as a
function of distance from the star.  The correlation between $q$ and
magnetic field strength is valid for all radii, though $q$ is not
constant everywhere.  For the cases with $\beta \le 1.0$, $q$ reaches
a maximum inside $r = 5R_*$, then decreases for increasing $r$.  This
can be understood if we note that the magnetic energy becomes much
weaker than total wind energy outside a few stellar radii for all
cases.  When $q$ is large, there are pressure gradients directed away
from the equator (i.e., the disk is compressed by the field, evident
in Fig.\ \ref{datamodel_fig}).  As the wind moves outward to the
region where the wind energy dominates, it can expand away from the
equator, causing a decrease in $q$ (the densities in Fig.\
\ref{angqtys_fig} are plotted near where the maximum contrast occurs
in all cases).  Here, of course, we have ignored radiative cooling in
the disk, which may be unrealistic.

The last two columns of Table \ref{results_tab} summarize the
simulation results of Figure \ref{contdens_fig}.  The maximum value of
$q$ for each $\beta$ and the values at $40R_*$ are listed.  For the
$\beta = 0.1$ and 0.2 cases, we can only place lower limits on the
maximum $q$ because the resulting disk is unresolved in our
simulations near the star (as discussed above).

\begin{figure}[t]
\plotone{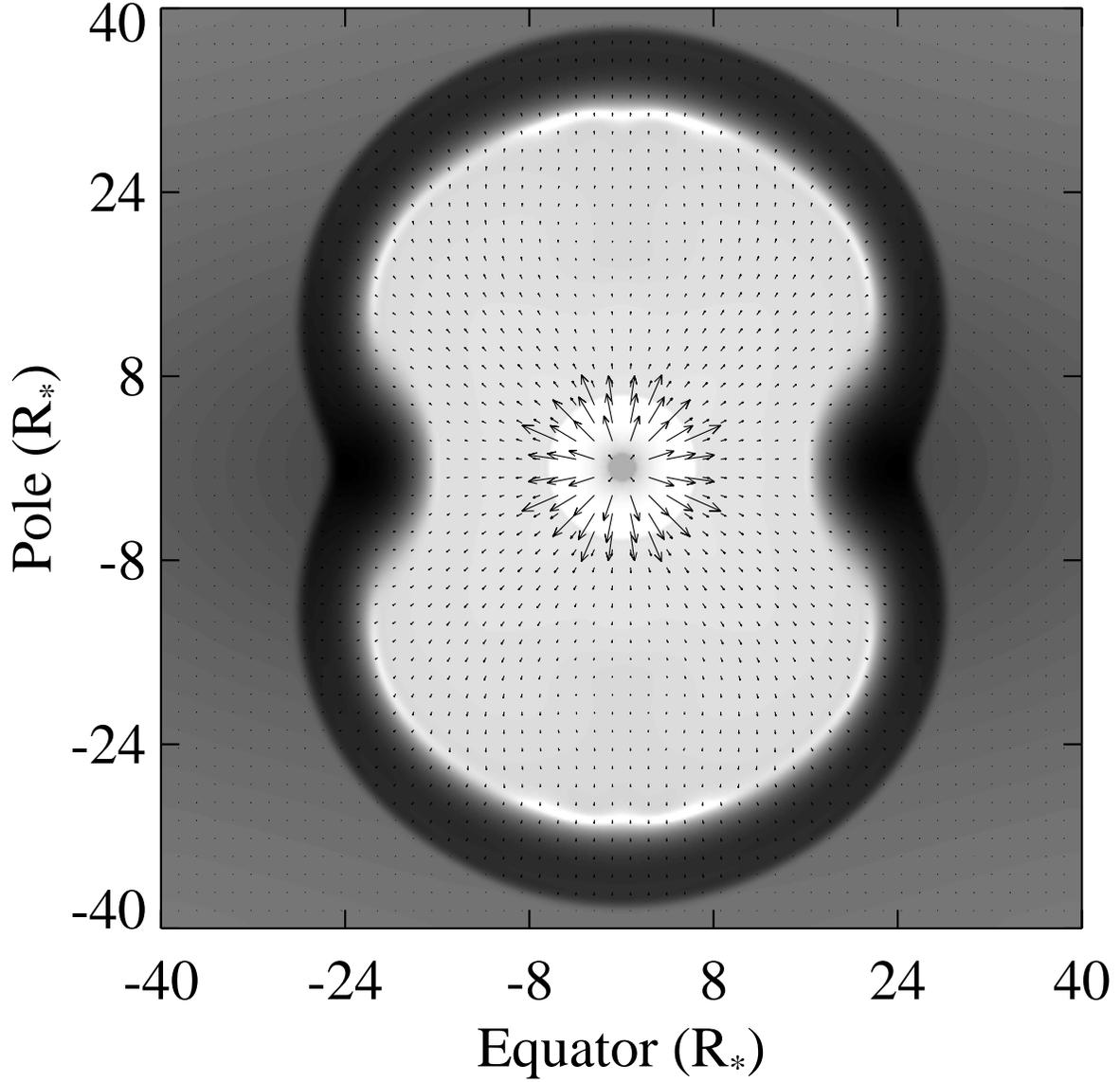}

\caption{An isotropic fast wind collides with an axisymmetric slow
wind.  The slow wind was produced by the mechanism discussed in the
text using a 2.7 Gauss ($\beta = 0.5$) dipole field.  Shown is $\log
n$ (cm $^{-3}$) grey-scale with $n \le 10^{7.5}$ as white and $n \ge
10^{11}$ as black.  Also shown are velocity vectors with the maximum
length corresponding to 1200 km s$^{-1}$.
\label{peanut_fig}}

\end{figure}

In general, density contrasts of roughly less than 2 in the slow AGB
wind may lead to elliptical PNe, while larger contrasts probably
produce bipolars \citep{f99}.  However, the density contrast is not
solely responsible for determining the shape of PNe, and the
distribution of mass is also important.  In order to superficially
test the inertial properties of the slow outflows produced in our
simulations, we ran another simulation that included a fast wind.  For
the initial conditions of this simulation, we used the end result of
the $\beta = 0.5$ simulation (see the bottom middle panel of Fig.\
\ref{denscont_fig}).  We then adjusted the pressure (increased by a
factor of 10) and density (decreased by a factor of 1000) on the
surface of the star so that a fast ($v \sim 1000$ km s${^-1}$),
tenuous ($\dot{\rm M} \sim 6 \times 10^{-7} \; {\rm M}_\odot$
yr$^{-1}$) wind blew from the star and collided with the slow-moving
disk, producing a shocked, wind-blown bubble.  On the surface of the
fast wind producing star, $\beta$ = 5.0, so the fast wind is affected
very little by the magnetic field on the surface.  Figure
\ref{peanut_fig} shows the density and velocity vectors on the
simulation grid 2 years after the onset of the fast wind.  The darkest
region in the figure corresponds to shocked, slow-wind material that
has been swept up in the fast wind.  While this simulation is too
simple (e.g., radiative cooling is ignored) to compare to real
objects, Figure \ref{peanut_fig} demonstrates that disks formed by
outflows from stars with dipole fields can be dynamically important
for shaping PNe.

\section{Discussion \label{discussion}}

We have presented a simple mechanism for producing an equatorially
concentrated wind from an isolated star.  The wind is isotropic in the
absence of magnetic effects.  In the presence of a dipole magnetic
field, however, the self-consistent interaction between the wind
plasma and the magnetic field causes the wind to become denser along
the magnetic equator and less dense near the poles.  The outflow model
is valid for very slowly rotating stars.  In fact, the model itself
requires no rotation, other than that necessary for generating the
magnetic field.

In the steady-state, the magnetic field retains much of its dipole
shape near the star, but becomes mostly radial outside a few stellar
radii and is oppositely directed above and below the magnetic equator.
This resembles the magnetic field configuration expected in the solar
wind \citep{an69,ws93,p91} with two major exceptions: 1) Solar
rotation adds an azimuthal (toroidal) component to otherwise radial
field lines, and 2) higher-order (though only important very near the
sun) and time-dependent magnetic fields \citep{bea98} are important on
the sun and further complicate the magnetic structure.

Our MHD simulations suggest that large equator-to-pole density
contrasts can be produced near an AGB star with a dipole magnetic
field strength of only a few Gauss on the surface.  Stronger fields
produce larger contrasts.  In addition to being denser, the wind near
the star is somewhat slower on the equator than along the poles.

As a simple test of validity, this model can be applied to the sun.
If we apply the argument given in \S \ref{siminit} to the solar wind
($\dot{\rm M} \sim 10^{-13} \; {\rm M}_\odot {\rm yr}^{-1}$, $v_{\rm
f} \sim 400$ km s$^{-1}$, $R_\odot \sim 7 \times 10^{10}$ cm), we find
that the wind and magnetic energy densities are comparable for a
magnetic field strength of 0.23 Gauss on the surface of the sun.  This
implies that the global average dipole field on the sun (often 1-2
Gauss) may have a significant effect on the isotropy of the solar wind
via the mechanism discussed here.  This is not surprising.
\citet{mestel68} and \citet{pk71}, for example, demonstrated that the
solar dipole is strong enough to retain closed field loops (see \S
\ref{model}) in the equatorial region.  The models of 
\citet{washimiea87} and \citet{kg99} indicated that such a field can
lead to an equatorial density enhancement in the solar wind.  On the
observational front, Ulysses measurements \citep{gea96} indicate the
solar wind is denser within 20 degrees of the solar equator than at
high latitudes by a factor of roughly 2.  We should note that the
solar wind is very complex, and the equatorial density enhancement in
the solar wind may have other or multiple origins.

According to the currently popular paradigm \citep{b87}, PNe may be
the result of a fast stellar wind colliding with a slower, denser
wind, producing an expanding shock front.  If the slow wind is
produced by the mechanism presented here, it will be denser and slower
in the equatorial plane.  These properties define an aspherical
inertial barrier (a nozzle) that would allow the PN to expand more
freely along the poles.  The extremity of this effect varies with
the strength of the star's dipole magnetic field while it is near the
end of its slow wind-producing phase.  Spherical PNe should result from
stars that had fields weaker than $\sim 0.5$ Gauss on the surface,
while stronger fields might lead to elliptical and bipolar PNe shapes.

We propose that sometime prior to the formation of the bright cores of
PNe, dipole magnetic fields are generated (via a dynamo) or exposed
(as winds remove the outer layers) on the surface of the star.  The
wind and field construct a toroidal circumstellar environment (or
outflowing disk) as described above.  Even if the subsequent fast
winds become isotropic, this disk can collimate the outflow until
instabilities or ablation destroy the disk.  We shall explore this
further in subsequent papers.  If the magnetic field is systematically
stronger for more massive stars, the observed correlation between
bipolarity and large progenitor mass \citep{zg88,cs95} is nicely
explained.

The development of the disk is essentially instantaneous with the
emergence of the magnetic field.  Thus, if the magnetic field were to
appear abruptly, one might expect mass loss to be initially isotropic
in character, followed by an abrupt change in geometry.  There is
observational evidence for such a ``mode change'' in AGB winds for
several objects \citep{hea97,th00,bw00}.  How such a change in the
magnetic field could occur in an AGB star is outside the scope of this
paper, since it implies a significant evolutionary change in the star.

The simplicity of the mechanism presented here makes it appealing, but
real stars are much more complicated.  More work is necessary to apply
the model to real stars.  For example, AGB winds are not likely to be
primarily pressure-driven as we have considered here.  A more
realistic treatment of the model should include the influence of
radiation pressure on dust grains \citep{hea94,sea97}, radiative
cooling, simulations in 3-D, the effects of higher-order magnetic
fields and fields with other symmetries (e.g.\ toroidal fields
produced by a dynamo or rotation), and the effects of stellar rotation
\citep{bc93,ws93}.  Also, the fact that our model reaches a steady-state
results from our assumption that the surface properties do not change.
In real AGB stars, pulsations may affect the wind (though if they are
isotropic, they can't effect the isotropy of the flow).  In the
presence of magnetic fields, convection and differential rotation
probably leads to more dynamic processes such as flares, coronal mass
ejections, spots \citep{frank95,soker98b}, magnetic cycles
\citep{soker00}, etc., as occur on the sun.

\begin{figure}[t]
\plotone{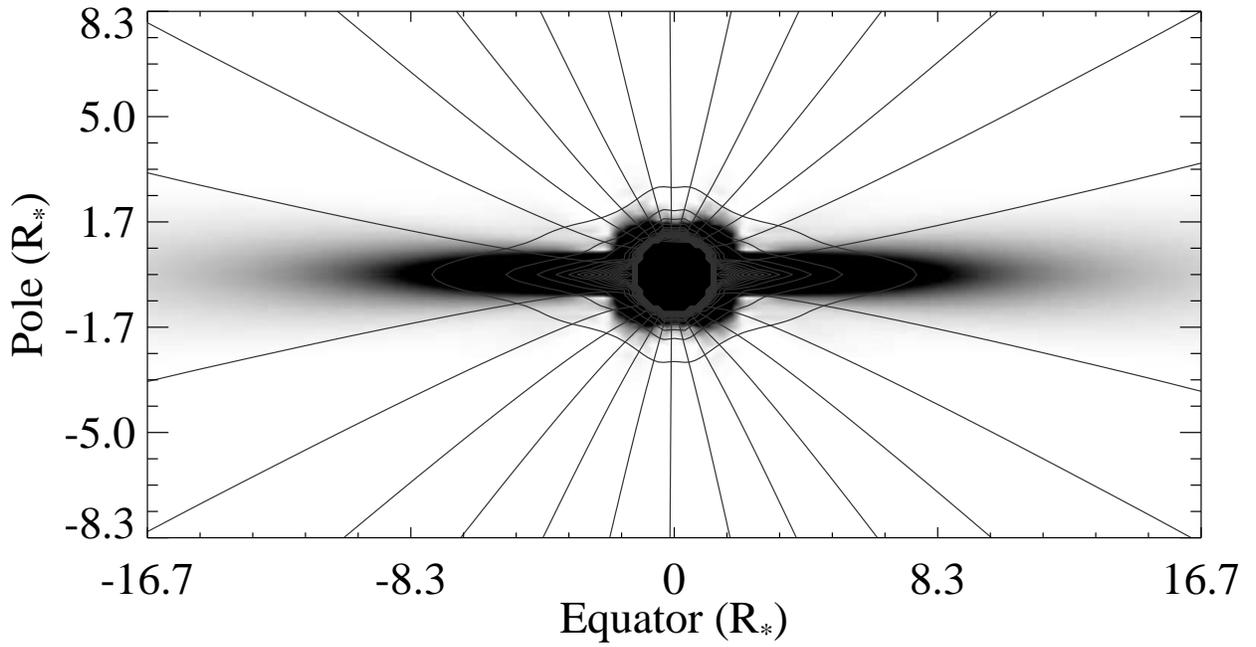}

\caption{Shown is an image of the absolute value of the azimuthal
current density for the $\beta = 0.2$ case after it has reached
steady-state.  Only the inner $200 \times 100$ grid points are shown.
Black corresponds to $\mbox{\boldmath $J$} \ge 1.5 \times 10^{-4} \,
\mu$A m$^{-2}$ while $\mbox{\boldmath $J$} \le 7.3 \times 10^{-6} \,
\mu$A m$^{-2}$ is white.  Magnetic field lines and pressure contours
are shown.  The central region corresponding to the interior of the
star has been artificially blacked out. \label{datamodel_fig}}

\end{figure}

\begin{figure}[t]
\plotone{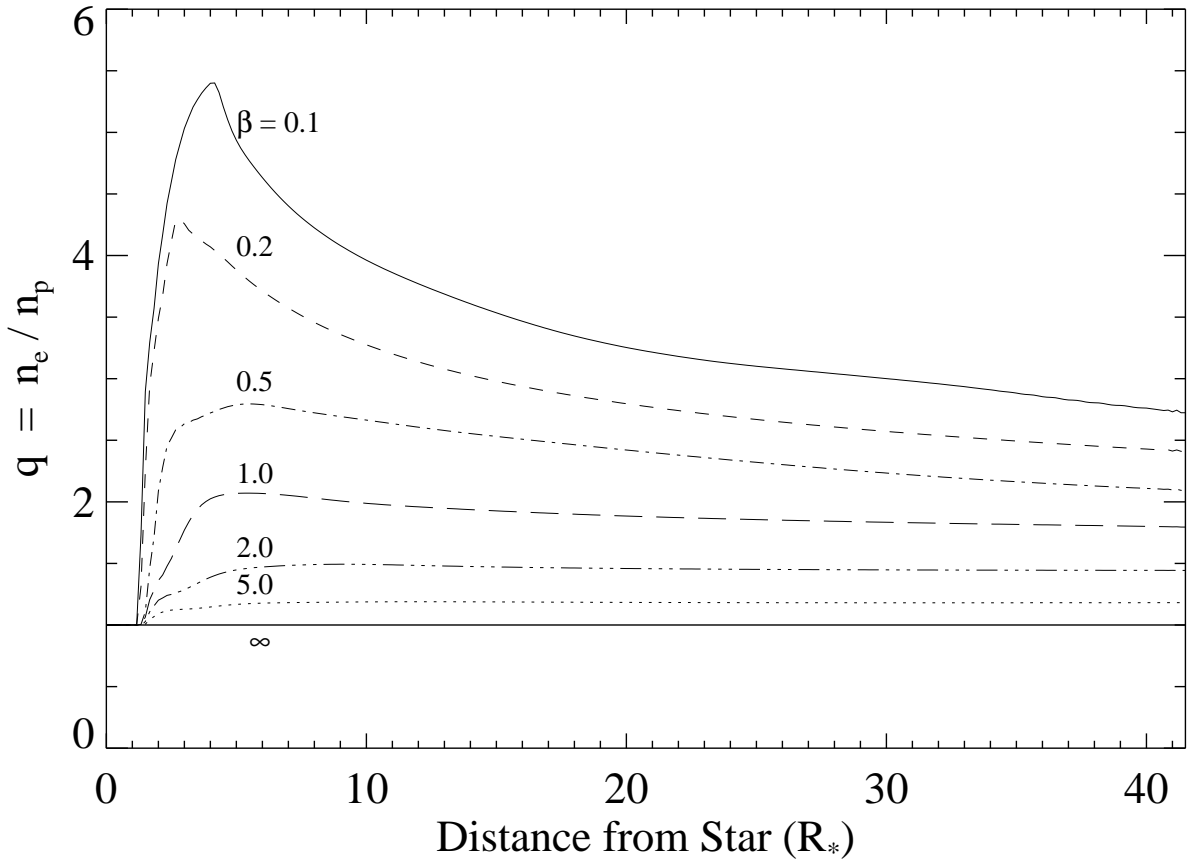}

\caption{The equator-to-pole density contrast, $q$, is shown as a function
of distance from the star for all cases.  The value of $\beta$ is
indicated next to the maximum $q$ of each case.
\label{contdens_fig}}

\end{figure}

In winds where pressure gradient forces are less important (e.g., for
radiation pressure-driven or radiatively cooled winds), the outflow
should behave differently.  For example, in our adiabatic simulations
the density contrast in the wind reached a peak and then decreased in
value further from the star (see Fig.\ \ref{contdens_fig}).  This is
probably due to pressure gradient forces directed away from the
equator (see Fig.\ \ref{datamodel_fig}).  If these forces are small,
the disk material may not expand as much (the disk will remain
flatter), leading to larger density contrasts than reported here.

While a more realistic approach than we have taken will affect the
details of the outflow, the qualitative model should apply to all
stars that have dipole magnetic fields of moderate strength (magnetic
energy comparable to the wind energy near the surface).

\acknowledgments

The authors would like to thank Adam Frank, Andrew Markiel, and Stacy
Palen for insightful discussion of many aspects of this work.  We are
also grateful to the referee, whose detailed comments led to
significant improvement of the paper.  This research was supported by
NSF grant AST-9729096.

\begin{deluxetable}{cccc}

\tablecaption{Simulation Results \label{results_tab}}
\tablewidth{0pt}
\tablehead{
\colhead{$\beta$} & \colhead{$\mbox{\boldmath $B$}_0$} & 
\colhead{$q$}   & \colhead{$q$}    \\
\colhead{}        & \colhead{(Gauss)}                  & 
\colhead{(max)}   & \colhead{($40R_*$)} 
}

\startdata
0.1 & 5.97 & $>$ 5.4 & 2.8 \\
0.2 & 4.22 & $>$ 4.3 & 2.5 \\
0.5 & 2.67 &     2.8 & 2.2 \\
1.0 & 1.89 &     2.1 & 1.8 \\
2.0 & 1.34 &     1.5 & 1.4 \\
5.0 & 0.84 &     1.2 & 1.2 \\
$\infty$ & 0.00 & 1.0 & 1.0
\enddata

\end{deluxetable}

\end{document}